\documentstyle[twocolumn,aps,prl]{revtex}
\begin{document}
\draft
\preprint{Draft copy --- not for distribution}
%
%
\wideabs{
%
%
\title{Infrared Studies of the Onset of Conductivity in Ultra-Thin Pb Films}
\author{P.F. Henning$^{1,2}$, C.C. Homes$^{1}$, S. Maslov$^{1}$,
G.L. Carr$^{1}$, D.N. Basov$^{2}$,
B. Nikoli\'{c}$^{3}$, and M. Strongin$^{1}$}
\address{
$^{1}$Department of Physics, Brookhaven National Laboratory, Upton, NY
11973}
\address{
$^{2}$Department of Physics, University of California at San Diego,
La Jolla, CA 92093}
\address{
$^{3}$Department of Physics, SUNY at Stony Brook, Stony Brook, NY
11794}
\date{\today}
\maketitle
%
%
\begin{abstract}
In this paper we report the first experimental measurement of 
the infrared conductivity of ultra-thin quenched-condensed 
Pb films. For dc sheet resistances such that $\omega \tau \ll 1$ the ac conductance increases with frequency but is 
in disagreement with the predictions of weak localization.  We attribute this 
behavior to the effects of an inhomogeneous granular structure 
of these films, which is manifested at the very small probing 
scale of infrared measurements. Our data are consistent 
with predictions of two-dimensional percolation theory.
\end{abstract}
\pacs{PACS numbers:  78.66.-w, 74.76.Db, 72.15.Rn, 64.60.Ak}
%
%
}
\narrowtext

Transport measurements in ultra-thin films have been a subject of
active interest over many years \cite{ultrathin}. These systems,
consisting of a thin layer of metal deposited onto a substrate held
at LHe temperatures, provide a relatively simple way to study the
interplay between localization, electron-electron interactions, and
superconductivity in disordered quasi-2D metals.
These experiments are in quantitative agreement with predictions
of localization theory \cite{loc} combined with the effects of
diffusion-enhanced electron-electron interactions \cite{aronov}.
The reason why these theories, developed for homogeneous materials,
work so well in the  case of granular, inhomogeneous films is that
the length scale at which electrons lose phase coherence in these
measurements is usually much larger than the characteristic size of
inhomogeneities (grains, percolation clusters, etc.) of the film.

Yet another way to modify the length scale
is by simply changing the probing frequency. This gives rise to
frequency dependence of the ac conductivity in the region
$\omega \tau \ll 1$, where the Drude theory predicts a plateau.
While the experimental data on the frequency dependence of conductivity
are virtually non-existent for ultra-thin quenched-condensed films,
they abound for thicker, more granular films, deposited onto a warm
substrate \cite{carr}.
In ac conductivity the frequency itself defines a characteristic
dephasing length scale $L_{\omega}=\sqrt{D/\omega}$, where
$D$ is a diffusion coefficient. In the frequency range
where $L_{\omega}$ is smaller than other dephasing length scales,
it enters into all localization and interaction formulas and gives
rise to frequency-dependent quantum corrections to the conductivity.
However, these quantum effects constitute only one source of
frequency dependence of the conductivity.  In the region where the
material is strongly inhomogeneous on the scale of $L_{\omega}$,
the frequency dependence of conductivity is dominated by purely
classical effects due to charge dynamics on a network of
capacitively and resistively-coupled clusters of grains.
The effective way to describe these effects theoretically is
provided by the framework of percolation theory \cite{frenchreview}.
In this theoretical approach the ac conductivity is shown to
increase with frequency.  Indeed, since capacitive coupling between
grains is proportional to frequency, grains become more and
more connected as the frequency is increased. It is also known
that purely quantum effects such as localization and
diffusion-enhanced interaction corrections become profoundly
modified on length scales where the material can no longer be treated
as homogeneous \cite{thouless}.  The presence of inhomogeneities is
known to change the functional dependence of these corrections on
the phase coherence length, while leaving the absolute magnitude
of the effect virtually unchanged.

In this letter we report the first measurement of conductivity at
infrared frequencies in ultra-thin films.  Films used in this experiment
were made {\it in situ} by evaporating Pb onto a Si(111) (sets 1 and
2) and glass (set 3) substrates, mounted in an optical cryostat, held
at 10~K.
Ag tabs, pre-deposited onto the substrate, were used to monitor the dc
resistance of the film.  Infrared transmission measurements from
500 to 5000~cm$^{-1}$ (set 1), and 2000 to 8000~cm$^{-1}$ (sets 2 and 3)
were made using a Bruker 113v spectrometer at the new high-brightness
U12IR beamline at the BNL's National Synchrotron Light Source.
The substrates were covered with a 5~\AA\ thick layer of Ge to promote
two-dimensional thin-film growth, rather than the agglomeration of the
deposited Pb in larger grains.  Even when using this traditional method,
which is known to promote the growth of more homogeneous films, where
the continuity has been measured at near a monolayer of deposited metal,
we only saw the beginnings of continuity near $8-20$~\AA\ of metal with
consistently thinner films using the glass substrates.  Films were
evaporated at pressures ranging from the low $10^{-8}$ to the mid
$10^{-9}$ Torr range.  The transmission spectra were obtained after
successive in-situ Pb depositions.  The dc resistances in set 1 on Si
range from $64\,{\rm M}\Omega/\Box$ at 17.4~\AA\ average thickness to
$543\,\Omega/\Box$ at 70~\AA .  The 70~\AA\ sample was then annealed
twice, first to 80~K, and then to 300~K.  As a result its resistance
at 10~K became $166\,\Omega/\Box$ after the first annealing, and
100 $\Omega/\Box$ after the second annealing.
Films from set 2 (also on Si) are similar to set 1:
we have observed $R_{\Box}=20\,{\rm M}\Omega/\Box$ at 18~\AA\ and
$R_{\Box}=1000\,\Omega/\Box$ at 88~\AA . Finally, films from set 3,
deposited on a Ge-coated glass substrate, range from 13
to 200~\AA, while $R_{\Box}$ changes between $5.6\,{\rm M}\Omega$ and
$22.8\,\Omega$.

The transmission coefficient of a film deposited on the
substrate, measured relative to the transmission of the
substrate itself, is related to real and imaginary
parts of the sheet conductance of the film as \cite{tinkhambook}
\begin{equation}
  {\rm T}(\omega) = {1\over {[1 + Z_0 \sigma^\prime_{\Box}(\omega)/
  {(n+1)}]^2 +
  (Z_0 \sigma^{\prime\prime}_{\Box}(\omega)/(n+1))^2} }.
\label{full_tinkham}
\end{equation}
Here $Z_0=377\,\Omega$ is the impedance of free space, $n$ is the
index of refraction of the substrate, equal to $n_{Si}=3.315$ for silicon
and $n_{G}=1.44$ for glass, and  $\sigma^\prime_{\Box}(\omega)$ (sometimes
called G), $\sigma^{\prime\prime}_{\Box}(\omega)$ are the real and
imaginary parts of the sheet conductance of the film. Almost
everywhere  in our experiments  $\sigma^\prime_{\Box}(\omega),
\sigma^{\prime\prime}_{\Box}(\omega) \ll (n+1)/Z_0$.  In this case
the contribution of the imaginary part of conductance to the
transmission coefficient is negligible and Eq.~(\ref{full_tinkham}) can
be approximately replaced by ${\rm T(\omega) \simeq \left[ 1 +
{Z_0 \sigma'_{\Box}(\omega)/(n+1)}\right]^{-2}}$.  Even for our thickest films, where 
$\sigma^{\prime\prime}_{\Box}(\omega) \approx (n+1)/Z_0$, the error in calculating $\sigma^\prime_{\Box}(\omega)$ in this way is less than ${\rm 10\%}$ over our frequency range.  
Throughout the manuscript we will use this approximation to extract the
real part of the sheet conductance of the film from its transmission
coefficient.  Only for our thickest films will we use Eq.~(\ref{full_tinkham})
to derive parameters of Drude fits.  

In Fig.~1 we plot the
frequency-dependent conductance, extracted from the transmission data
for the films from the set 3 with the help of the above approximation
to Eq.~(\ref{full_tinkham}).  The seven thickest films from this set
exhibit a characteristic Drude falloff at high frequencies.
For the rest of the films the conductivity systematically
increases with frequency throughout our frequency range. The inset in
Fig.~1 shows the average ac conductance as well as the dc sheet conductance
for set 3 as a function of thickness. Note the curves start to significantly
deviate from each other at around 50~\AA.

In order to fit the conductance of our thickest films with the Drude
formula, one needs to use the untruncated Eq.~(\ref{full_tinkham}) for
the transmission coefficient.  Inserting the Drude expression
for the sheet conductance $\sigma_{\Box}(\omega)=\sigma_D/(1-i\omega \tau)$
directly into Eq.~\ref{full_tinkham} one gets ${\rm T}(\omega)/[1-{\rm T}(\omega)]=(1+\omega ^2 \tau^2)/
[(\sigma_{D}/\sigma_{0})^2+2\sigma_{D}/\sigma_{0}]$, where $\sigma_0=(n+1)/Z_0$.
Therefore, the transmission data, which are consistent with the Drude
formula can be fitted with a straight line, when ${\rm T}(\omega)/[1-
{\rm T}(\omega)]$ is plotted as a function of $\omega^2$.  In Fig.~2 we
plot the 7 thickest films from the set 3 in this way.  The knowledge of
the average thickness of our films along with parameters of the Drude
formula enables us to calculate the plasma frequencies in our films.
They are shown in the inset of  Fig.~2 as a function of $1/\sigma_D$ ---
the dc sheet resistance in Drude formula, which itself was extracted
from our Drude fit.  These results are in excellent agreement with
the experimentally determined lead plasma frequency of
$\omega_p = 59\,400$~cm$^{-1}$ \cite{plasmaexp}.
In the remainder of the manuscript we discuss possible interpretations
of the increase of the conductance with frequency, which we observe in
our thinner films.

One mechanism which is known to cause a frequency dependence of
conductivity within a Drude plateau $(\omega \tau \ll 1)$ is
purely quantum mechanical in origin.  The conductivity is known to
be reduced due to increased back scattering of phase-coherent
electrons (so called weak localization (WL) \cite{loc}), as well as
diffusion enhanced electron-electron interactions (EEI) \cite{aronov}.
The magnitude of this reduction depends on the length scale over
which an electron maintains its phase coherence. In the absence of
external magnetic field or ac electric field this length is determined
by the temperature.  It is given by the inelastic scattering length
$l_{in}(T)=\sqrt{D \tau_{in}(T)}$ for WL, and the thermal coherence
length  $L_{T}=\sqrt{\hbar D/kT}$ for EEI. Here $D$ is the diffusion
coefficient of the electron related to its dc conductivity by the
Einstein formula $\sigma = e^2 (dN/d\mu)_{E_F} D$, and $\tau_{in}(T)$
is the temperature dependent inelastic scattering (dephasing)
time.  In the presence of the ac electric field the
diffusive motion of an electron is restricted to a spatial region
of size $L_{\omega}=\sqrt{D/\omega}$.
If this length scale turns out to be shorter than the corresponding
dc length scale, it is $L_{\omega}$ which enters in all WL and EEI
formulas.

The question of effective dimensionality of the quasi-2D sample is
decided by comparing $L_{\omega}$ to the film thickness $d$.
The frequency dependent WL corrections to the sheet conductance of the
film are given by
$\Delta \sigma_{\Box}^{2D} (\omega) =
{e^2 \over 2 \pi^2 \hbar} \ln \omega \tau$
in the 2D limit ($d<L_{\omega}$) \cite{loc}, and
$\Delta \sigma_{\Box}^{3D} (\omega)=
{\sqrt{2} e^2 \over 4 \pi^2 \hbar} d \sqrt{\omega \over D} \label{3D_loc}$
in the 3D limit ($d>L_{\omega}$) \cite{gorkov}.
At the lower end of our frequency range $\omega=500$~cm$^{-1}$
for a realistic value of $D=5$~cm$^2$/s we can estimate
$L_{\omega}\leq 20 {\rm \AA}\leq d$.  Therefore, for our films one
should use the formulas of three-dimensional localization
theory.  The frequency-dependent sheet conductance in most of our
films is consistent with the $\sqrt{\omega}$ dependence of 3D WL.
However, we believe that in order to explain the frequency dependence
of our experimental data one needs to look for yet another mechanism,
supplementing that due to weak localization and electron-electron
interactions.  The problems with ascribing the observed frequency
dependence of conductivity solely to WL and EEI effects are:
(i) the dependence of the slope of the conductivity vs $\sqrt{\omega}$
on the thickness of the film and the dc sheet conductance, which determines
the diffusion coefficient $D$,  does not agree with predictions of
the 3D localization.
(ii) the weak localization theory is only supposed to work in the limit
where its corrections are much smaller than the dc conductivity. In our
experimental data we don't see any change of behavior as the corrections
to conductivity become bigger than the dc conductivity.
In fact the $\sqrt{\omega}$ fit works very well and gives roughly the
same slope even for films with dc sheet resistance of $\approx 100$~k$\Omega$,
while the ac sheet resistance is only $\approx 1\,{\rm k}\Omega$.
Furthermore, the 3D-localization theory predicts that the
$\sqrt{\omega}$ dependence of  weak localization theory should be
replaced by $\omega^{(d-1)/d}=\omega^{1/3}$ dependence at or near the
3D metal-insulator transition \cite{vollhardt}.  In our experimental
data we see no evidence for such a crossover.

There exists yet another, purely classical effect that gives rise
to the frequency dependence of the conductivity. It is relevant in strongly
inhomogeneous, granular films.  There is ample experimental evidence that
even ultra-thin quenched-condensed films have a microscopic granular
structure \cite{vallesAu,kagawa}.  In order to describe the ac response of a film with such a granular microstructure one needs to know the geometry and conductivity
of individual grains as well as the resistive and capacitive couplings
between grains.  The disorder, which is inevitably present in
the placement of individual grains, makes this problem even more
complicated.  However, there exist two very successful approaches
to the analytical treatment of such systems.  One of them, known as
the effective-medium theory \cite{EMT}, can be viewed as a mean-field
version of a more refined approach, based on scaling near the
percolation transition.  The EMT takes into account only
concentrations of metallic grains and the voids between the
grains, disregarding any spatial correlations.  A more refined approach takes
into account the geometrical properties of the mixture of metallic
grains and voids.  The insulator-to-metal transition in this approach
is nothing else but the percolation transition, in which metallic
grains first form a macroscopic connected path at a certain critical
average thickness $d_c$ of the film.
The dc conductivity above the transition point scales as $(d-d_c)^t$,
where  $t=1.3$ in 2D and $t=1.9$ in 3D \cite{frenchreview}. Just below
the percolation transition the dielectric constant of the medium
diverges as $\epsilon (d) \sim (d_c-d)^{-s}$, where
$s=1.3$ in 2D and $s=0.7$ in 3D.  The diverging dielectric
constant is manifested as the imaginary part of the ac conductivity
$\sigma(\omega) \sim -i\omega (f_c-f)^{-s}$.
In general the complex ac conductivity of the metal-dielectric (void)
mixture close to the percolation transition is known \cite{frenchreview}
to have the following scaling form:
\begin{equation}
  \sigma(\omega,d)=|d-d_c|^{t} F_{\pm}(-i\omega |d-d_c|^{-(t+s)}).
\label{perc_scaling}
\end{equation}
Here $F_{+}(x)$ and $F_{-}(x)$ are scaling functions
above and below the transition point correspondingly.
Note that this scaling form correctly reproduces the scaling of
the dc conductivity above the transition and the
divergence of the dielectric constant below the transition provided that
$F_{+}(x)=F_{+}^{(0)}+F_{+}^{(1)}x+F_{+}^{(2)}x^2+ \ldots$, while
$F_{-}(x)=F_{-}^{(1)}x+F_{-}^{(2)}x^2+ \ldots$  One should mention
that the predictions of the EMT can also be written in this scaling
form with mean-field values of the exponents $t=s=1$, and scaling
functions $F{\pm}(x)=(\sqrt{D^2+4(D-1)x} \pm D)/(2(D-1))$, where
$D$ is the spatial dimension.

Since the metallic grains in our films form not more than two layers,
our data should be interpreted in terms of the two-dimensional
percolation theory.  In two dimensions $t=s=1.3$ \cite{frenchreview},
and according to Eq.~(\ref{perc_scaling}) the ac conductivity
{\it precisely} at the transition point $d=d_c$ is given by
$\sigma(\omega, d_c)=A (i \omega/\omega_0)^{t/(t+s)}=A (i 
\omega/\omega_0)^{1/2}$.
This prediction is in agreement with our experimental data.  In Fig.~3
we attempt the rescaling of our data according to Eq.~(\ref{perc_scaling}).
The critical thickness $d_c$ is determined as the point where the ac
conductivity divided by $\sqrt{\omega}$ is frequency independent.
Of course, the experimental uncertainty in our data points does not
allow us to determine which exponents $t$ and $s$ provide the best
data collapse.  However, as we can see from Fig.~3 our data are
{\it consistent} with the scaling form of the 2D percolation theory.
Finally, we use Fig.~3 to estimate basic parameters such as typical
resistance $R$ of an individual grain and typical capacitance $C$
between nearest neighboring grains.  From the limiting value of
$\sigma (\omega,d) (d_c/|d-d_c|)^{1.3}$ at small values of
the scaling variable $x=\omega (d_c/|d-d_c|)^{2.6}$ for $d>d_c$,
one estimates the conductance of an individual grain to be of order of
$R \sim 1000\,\Omega$.  In the simplest $RC$ model, where the fraction
of the bonds of the square lattice are occupied by resistors of
resistance $R$, while the rest of the bonds are capacitors with
capacitance $C$, the ac conductivity exactly at the percolation
threshold is given by $A/R (i \omega RC)^{1/2}$, where $A$
is a constant of order of one.  Therefore, the slope
$\partial \sigma/\partial \sqrt{\omega}$ in our system should be
of the same order of magnitude as $\sqrt{RC}/R$. This gives
$C \simeq 2.6 \times 10^{-19}$~F, which is in agreement with
a very rough estimate of the capacitance between two islands
$200\,{\rm \AA} \times 200\,{\rm \AA} \times 30\,{\rm \AA}$
separated by a vacuum gap of some $20\,{\rm \AA}$, giving
$C \simeq 2.7 \times 10^{-19}$~F.  This order of magnitude
estimate confirms the importance of taking into account
inter-island capacitive coupling when one interprets the
ac conductivity measured in our experiment. Indeed, $R=1000\,\Omega$,
and $C=3 \times 10^{-19}$~F define a characteristic frequency
$1/RC \simeq 17000$ cm$^{-1}$ comparable to our frequency range.

In summary, we have measured the conductivity of ultra-thin Pb films in the frequency range 500 to 8000~cm$^{-1}$.  The evolution of $\sigma(\omega )$ with DC sheet resistance is consistent with classical two-dimensional percolation theory in this range.  At lower probing frequencies, where $L_{\omega}$ becomes larger than the scale of inhomogeneities in these films, we expect that the effects of weak localization will become more prevalent.

We have benefited from fruitful discussions with P.B.~Allen,
A.M. Goldman, V.J. Emery, V.N. Muthukumar, Y. Imry, Z. Ovadyahu,
and M. Pollak. The work at Brookhaven was supported by the U.S.
Department of Energy, Division of Materials Sciences, under contract
no.~DE-AC02-98CH10886.  Support from NSF grants DMR-9875980 (D.N.B.) and DMR-9725037 (B.N.) is
acknowledged.  Research undertaken at NSLS is supported by the
U.S. DOE, Divisions of Materials and Chemical Sciences.

%
%

%
%
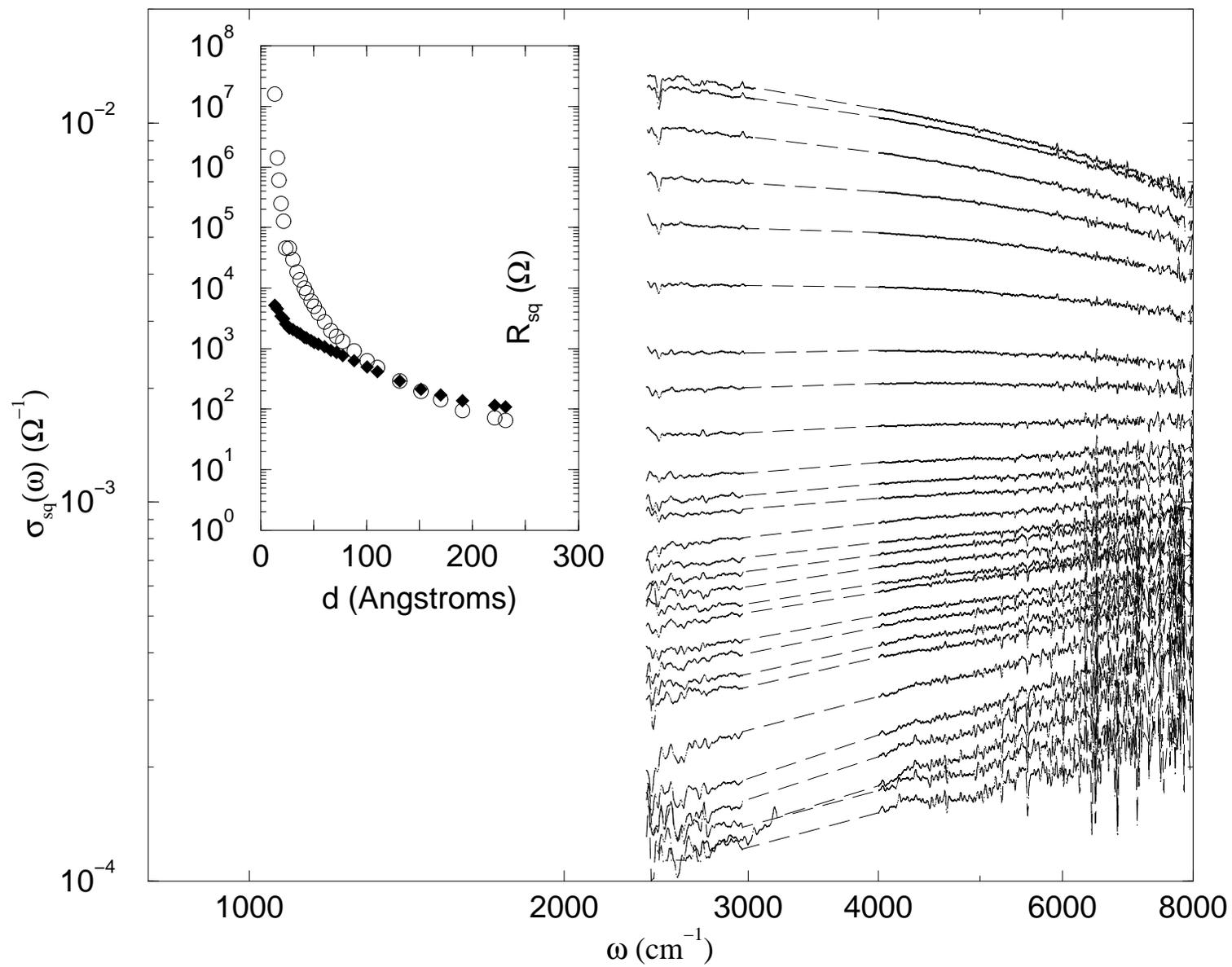
\begin{figure}
\caption{Sheet conductance vs frequency for set 3. The dashed lines plotted between 3000 and 4000~cm$^{-1}$ (where the glass substrate is opaque) are a guide to the eye.  The inset shows the
  inverse average conductance in our frequency range (filled symbols) and
  the dc sheet resistance (open symbols) as a function of the film thickness.}
\end{figure}

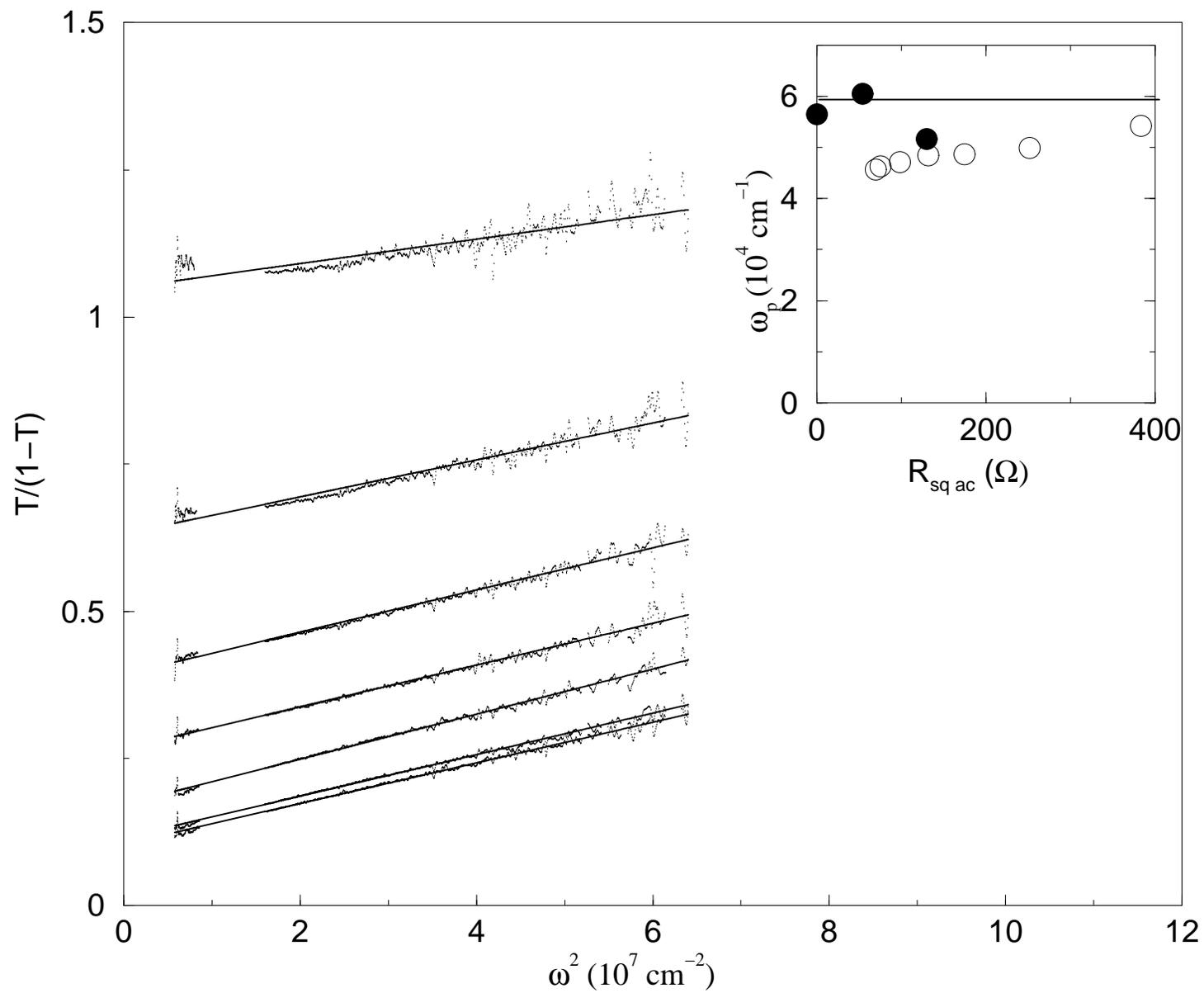
\begin{figure}
\caption{T/(1-T) plotted vs. $\omega^2$ for the 7 thickest films from set 3; The solid lines are Drude Model fits, as described in the text.  The inset shows the plasma frequency for
  these films (open symbols) and three quenched films from sets 1
  and 3 (filled symbols) as a function of the dc sheet resistance, with the
  flat line representing the plasma frequency of bulk lead obtained
  from Ref.~8.}
\end{figure}

\begin{figure}
\caption{The rescaling of conductivity data in set 3 according to
 predictions of the percolation theory. The value for $d_c$ is
 $34\,{\rm \AA}$.}
\end{figure}

%
%
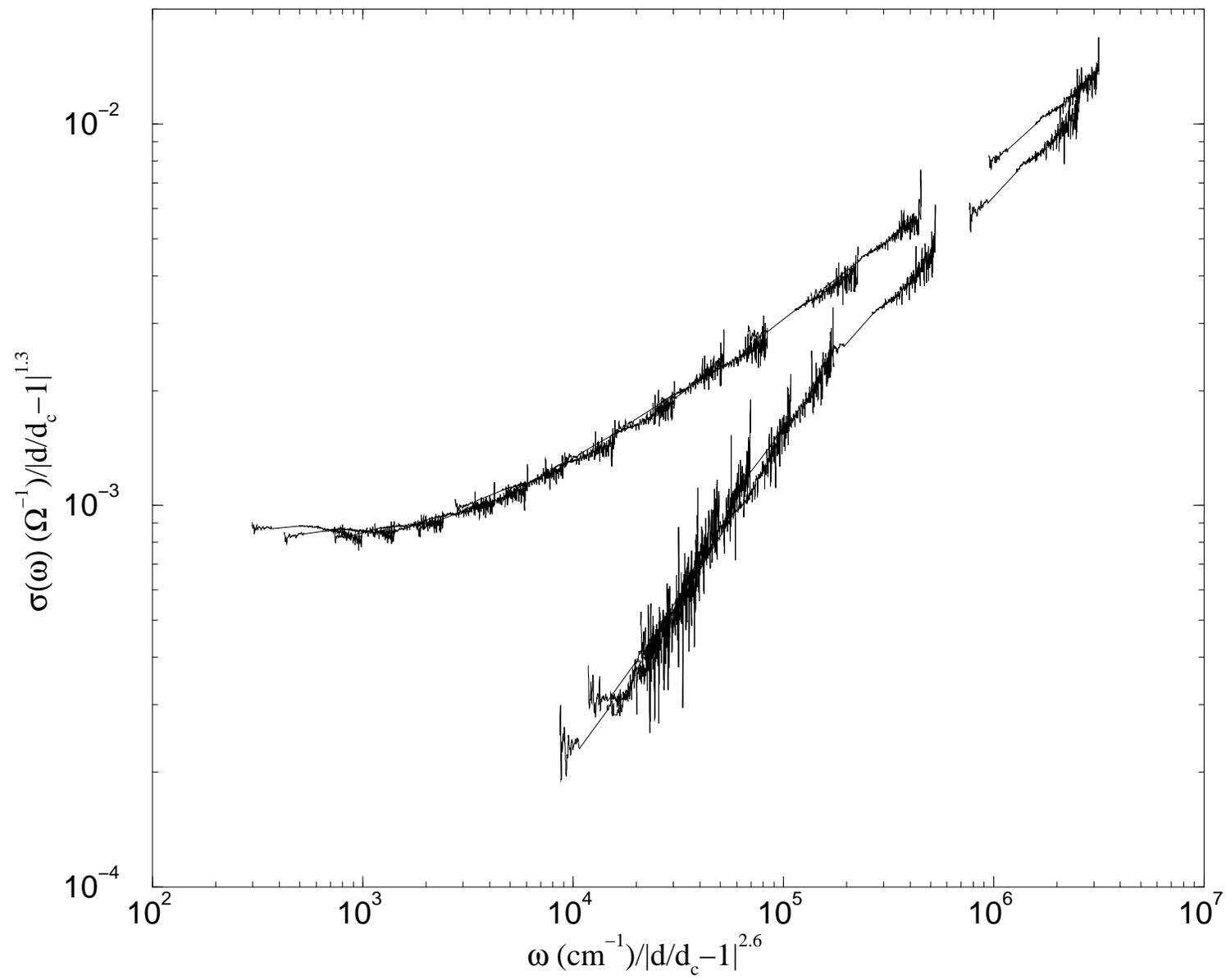
\end{document}